\documentclass[]{jfm_mod_noj} 

\RequirePackage{packages}
\RequirePackage{newcommands}

\newcommand{\srev}[1]{{\color{black} #1}}

\renewcommand{\pbut}{\pbu_\mathrm{tot}}
\newcommand{\oluzt}{\ol{u_{z,\mathrm{tot}}}}
\newcommand{\acpsi}{\acute{\psi}}

\ifCUPmtlplainloaded \else
\checkfont{eurm10}
\iffontfound
\IfFileExists{upmath.sty}
{\typeout{^^JFound AMS Euler Roman fonts on the system,
using the 'upmath' package.^^J}%
\usepackage{upmath}}
{\typeout{^^JFound AMS Euler Roman fonts on the system, but you
dont seem to have the}%
\typeout{'upmath' package installed. JFM.cls can take advantage
of these fonts,^^Jif you use 'upmath' package.^^J}%
\providecommand\upi{\upi}%
}
\else
\providecommand\upi{\upi}%
\fi
\fi

\ifCUPmtlplainloaded \else
\checkfont{msam10}
\iffontfound
\IfFileExists{amssymb.sty}
{\typeout{^^JFound AMS Symbol fonts on the system, using the
'amssymb' package.^^J}%
\usepackage{amssymb}%
  \let\leq=\leqslant
  \let\geq=\geqslant
}{}
\fi
\fi
 
\ifCUPmtlplainloaded \else
\IfFileExists{amsbsy.sty}
{\typeout{^^JFound the 'amsbsy' package on the system, using it.^^J}%
\usepackage{amsbsy}}
{\providecommand\boldsymbol[1]{\mbox{\boldmath $##1$}}}
\fi

\title[]{
Designing vortices in pipe flow with topography--driven Langmuir circulation 
}

\author[A.~H. Akselsen, L.~Chan and S.~\AA. Ellingsen]{
 Simen \AA. Ellingsen$^1$\thanks{Email address for correspondence: simen.a.ellingsen@ntnu.no. All authors contributed equally and are to be considered joint first authors.}, 
Andreas H.\ Akselsen$^{1,2}$
and Leon Chan$^3$}

\affiliation{$^1$Department of Energy and Process Engineering, Norwegian University of Science and Technology, N-7491 Trondheim, Norway.
\par$^2$ SINTEF Ocean, Marinteknisk senter, N-7052 Trondheim, Norway.
\par$^3$ Department of Mechanical Engineering, The University of Melbourne, Victoria 3010, Australia}
\date{\today}           

\begin{document}

\maketitle

\begin{abstract}

\srev{
We present direct numerical simulation of a mechanism for creating longitudinal vortices in pipe flow, compared with a simple model theory. By furnishing the pipe wall with a pattern of crossing waves secondary flow in the form of spanwise vortex pairs is created. The mechanism `CL1' is kinematic and known from oceanography as a driver of Langmuir circulation. 
CL1 is strongest when the `wall wave' vectors make an accute angle with the axis, $\varphi=10^\circ$ - $20^\circ$ (a `contracted eggcarton'), changes sign near $45^\circ$ and is weak and opposite beyond this angle. 
A competing, dynamic mechanism driving secondary flow in the opposite sense is also observed created by the azimuthally varying friction. Whereas at smaller angles `CL1' prevails, the dynamic effect dominates when $\varphi\gtrsim 45^\circ$ reversing the flow. 
Curiously, circulation strength is a faster--than--linearly increasing function of Reynolds number for the contracted case.

We explore an analogy with Prandtl's secondary motion of the second kind in turbulence. A transport equation for average streamwise vorticity is derived, and we analyse it for three different crossing angles, $\varphi=18.6^\circ, 45^\circ$ and $60^\circ$. Mean-vorticity production is organised in a ring-like structure with the two rings contributing to rotating flow in opposite senses. For the larger $\varphi$ the inner ring decides the main swirling motion, whereas for $\varphi=18.6^\circ$ outer--ring production dominates. For the larger angles the outer ring is mainly driven by advection of vorticity and the inner by deformation (stretching) whereas for  $\varphi=18.6^\circ$ both contribute approximately equally to production in the outer ring. 

}

\end{abstract}


 
\newcommand{\ve}{\varepsilon}
\renewcommand{\p}{ }
\newcommand{\order}{\mathcal{O}}
\renewcommand{\mnil}{m}
\renewcommand{\knil}{k}

\newcommand{\up}{\breve{p}}
\newcommand{\uq}{\breve{q}}
\newcommand{\uu}{\breve{u}}
\newcommand{\ubu}{\breve{\bu}}
\newcommand{\upsi}{\breve{\psi}}


\section{Introduction}
\label{sec:introduction}

Secondary mean motion in the form of coherent streamwise vortices has often been employed to favourably manipulate transitional and turbulent pipe flow and wall-bounded flows. Approaches to flow control based directly or indirectly on the creation of streamwise vortices in wall-bounded flow, are many and varied, including both active and passive schemes.  Introducing streamwise vortices with carefully designed roughness elements was shown by \citet{fransson06} to delay transition to turbulence, and actively introducing vortices was shown to favourably redistribute turbulence \citep{willis_2010_optimally_amplified_streak_drag_rediction_PhysRev} or remove it altogether \citep{kuhnen_2018_destabilizing_of_turbulence}. 
Active methods implemented experimentally include cross-flow jets \citep{iuso02}, blowing and suction \citep{segawa_2007_drag_reduction_blowing_and_suction,lieu10} and individually rotating wall segments \citep{auteri10}. A common denominator in all these approaches is the search for ways to reduce boundary layer skin friction. 

The use of specially designed wall roughness elements is a well established idea for the manipulation of boundary layer flows. Vortical secondary flow has been shown in a number of studies to result from spanwise intermittent roughness patches  \citep{anderson15,willingham14} and streamwise aligned obstacles \citep{yang18,vanderwel15, kevin17, sirovich97}. \citet{anderson15} later demonstrated that these structures are related to Prandtl's secondary flow of the second kind, driven by spatial gradients in the Reynolds-stress
components.
\srev{
More recent studies show that intentionally imposed near wall streaks and hairpins can stabilise the overall flow regime and delay or
prevent transition into turbulence \citep{du00, cossu02, cossu04,fransson05, fransson06, pujals10a, pujals10b}.
}
Most directly related to the current study, \cite{chan15, chan18} studied pipe flow by way of direct numerical simulation (DNS) wherein an `egg carton' structured wall roughness was introduced composed of sine waves crossing at right angles, a special case of the geometry considered in the present paper. These authors also report secondary motion in the form of vortices in the time-averaged flow oriented streamwise. 

\srev{

\citet{bhaganagar04} considered wall-bounded turbulent flow with egg--carton type roughness from a crossing wave pattern, comparing it to a smooth wall. While secondary flows were not studied explicitly, varying the crossing angle and steepness of the waves was found to affect the the outer boundary layer even though roughness elements did not extend beyond the viscous sub-layer, an indication that coherent motions at a much larger scale were set in motion. A somewhat similar study of turbulent flow over a pyramidal pattern by \citet{hong11} showed a mechanism where roughness-size vortices were created then lifted into the bulk. The fact that their roughness was contained within the laminar sub--layer makes us conjecture that the mechanism studied by \citet{akselsen20} and herein has relevance for turbulent flows, particulary the debate whether and how the outer part of a boundary layer is affected by the detailed morphology of the wall roughness \citep{bhaganagar08, antonia10}. 

}

All of the above mentioned
secondary flows induced by wall topography or roughness, however, are driven by essentially dynamic mechanisms relying on gradients in viscous stress. In contrast, we here consider a passive mechanism for vortex generation which is of kinematic origin and a close analogy of a mechanism for Langmuir circulation, a phenomenon known from a traditionally disparate branch of fluid mechanics: Oceanograhic flow. 
Langmuir circulation is a motion in the form of long streamwise and evenly spaced vortices just beneath the surface of oceans or lakes. \citep{leibovich83}. The vortices are often clearly visible as `windrows' --- near-parallel lines of debris gathering in the downwelling regions between vortices. There are two principal mechanisms by which Langmuir circulation is created of which we make deliberate use of the one often referred to as `CL1' in honour of the pioneering theory of \citet{craik76}. The motion is driven by a resonant interaction between sub-surface shear currents and waves, both typically generated by the wind. It was suggested as a Langmuir flow mechanism by \cite{craik70} inspired by \cite{benney60}, and works by twisting spanwise vorticity already present in the ambient shear flow into the streamwise direction via the wave--induced Stokes drift; see \citep{leibovich83}. 
In our case the near--surface shear layer is replaced by boundary layer shear, and surface waves by a wavy wall.

To uncover the nature of the Langmuir vortices we consider only laminar flow. Their stability, prevalence and effects in turbulent pipe flow remains an open and potentially important question for the future, 
\srev{
yet our study does shed a modicum of light on that question. Upon averaging the Navier--Stokes equations over one spanwise period of our geometry, they obtain a form identical to the Reynolds--averaged Navier--Stokes equations in streamwise--uniform geometry but for the definition of the averaging operator. The analogue of Reynolds stress we refer to as undulation stress and results from a combination of inviscid pressure forces from the flow being guided by the curved walls, and viscous forces from spanwise--varying roughness. These two drive the Langmuir mechanism and a (for certain parameters) competing dynamic drag mechanism of mean vortical flow. The analogy is closely related to the double--averaging concept due to \citet{nikora07}, whereby temporal/ensemble averaging is supplemented by and spatial averages over volumes, areas or distances. We explore this concept further in Section \ref{sec:Prandtl}.

Creating vortices in laminar flow is of considerable interest in itself for the purpose of mixing in microfluidic channels. The use of imprinted wall features for passive mixing is a long--established method in microfluidic flow systems \citet{ward15}, for instance the use of oblique ridges to twist and fold the flow has been highly impactful \cite{stroock02}. 
Vortical motion can greatly enhance heat transfer, important e.g.\ for direct liquid cooling of high power density electronic devices; secondary flow (Dean vortices) generated by guiding fluid through wavy microchannels \citep[e.g.][]{sui10} is a popular method for efficient mixing with low pressure drop penalty. 

The mechanisms here considered are superficially similar to, but distinct from, several phenomena which have received attention in recent turbulence literature. A theory for an instability in Couette flow in a channel with periodically modulated walls in the streamwise directions was recently derived by \cite{hall20}, in turn related to one previously analysed by \citet{floryan02,floryan03,floryan15} and \cite{cabal02}. 
Unlike `CL1' this is an instability rather than a directly driven secondary flow, occuring beyond a critical Reynolds number depending on wall corrugations, and the geometry of these studies varies in the streamwise, but not spanwise directions. Several studies see streamwise streaks for purely spanwise boundary modulations \citep[e.g.,][]{colombini95,willingham14,anderson15, hwang18} whose relation to our study we discuss in 
Section \ref{sec:hmp}.
In simulation, \citet{schmid92} found that transition to turbulence was much accelerated through growth of streamwise vortices when a a pair of finite--amplitude oblique waves were initially imposed. The link to our work is not obvious, yet we note that the presently reported mechanism is due to interactions of pairs rather than triads of wave modes. Herringbone-type riblets have been demonstrated to reduce viscous drag in turbulent boundary layer flows \citep{walsh83} and, like other laterally inhomogeneous roughness geometries, also exhibit large secondary motion in the form of streamwise rolls \cite{kevin19}. The strong ejections due to fluid being forced upwards where the yawed riblets converge, however, sets this flow somewhat apart.

The outline of the paper is as follows. We begin in Section \ref{sec:theory} with a model theory for the Langmuir-type vortical motion, along with, in Section \ref{sec:results}, theoretical predictions pertinent to our numerical investigation, which follows in Section \ref{sec:sim_results}. A discussion of the analogy to Prandtl's second mechanism of secondary motion in turbulence follows in Section \ref{sec:Prandtl} before Conclusions. Some additional theory of initial vortex growth is found in an appendix, and a collection of results of all simulated cases are provided as online supplementary material. 
}


\section{Model theory for creation of Langmuir--type vortices}
\label{sec:theory}

\begin{figure}%
\includegraphics[width=\textwidth]{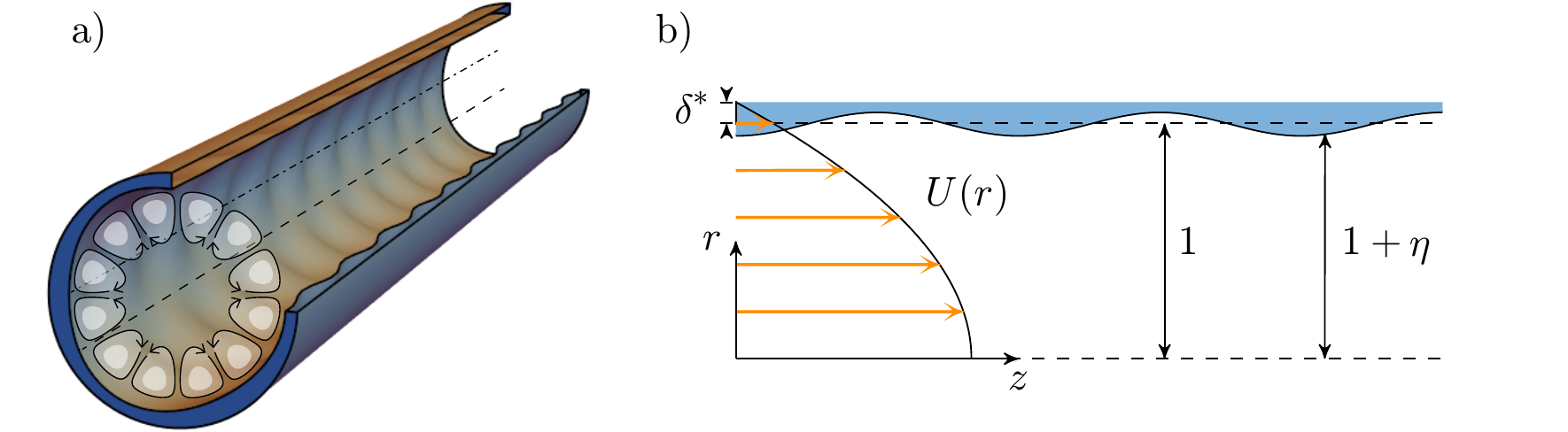} %
\caption{a) Pipe geometry for $m=3, \kappa=5$; crestlines (dash-dot) and saddle--point lines (dash) are shown; b) Geometry and parameters used in section \ref{sec:theory}.}
\label{fig:schematic}%
\end{figure}

We construct a simplified theory of perturbations, revealing the origin of Langmuir-type vortices. Our geometry is illustrated in Figure~\ref{fig:schematic}a 
\srev{
consisting of an infinitely long circular pipe whose walls are augmented by the addition of a pattern of crossing waves.
The steepness of these `wall waves' measured in the streamwise direction is presumed to be small: $\ve=ka\ll 1$ where $a$ is the waves' amplitude and $k$ their streamwise wave number. The amplitude is also presumed much smaller than the radius, $a\ll 1$. We proceed in increasing orders of $a$ assuming a basic flow of parabolic Poiseuille form with centerline velocity $U_0$. 
}

We first non-dimensionalise using pipe radius $R$ and $U_0$ of the basic flow:
\begin{align}
(r,z)&\mapsto (r,z)\,R,
&
k&\mapsto k\,R^{-1},
&
t & \mapsto t R/U_0,
&
\p p &\mapsto \p p\, \rho U_0^2,
&
 \pbut &\mapsto  \pbut\,U_0,
&&
\end{align}
where $\pbut$ here denotes any measure of fluid velocity, 
$\rho$ is the fluid density,
and $\p p$ the pressure perturbation 
on top of the 
constant pressure gradient driving the mean flow. We ignore gravity throughout. The bounding surface is now perturbed
slightly and is found at $r=1+\p\eta(z,\theta)$ where $|\p\eta|\sim a\ll 1$.

We write the resulting three-dimensional velocity field as 
$\pbu(r,\theta,z,t)=\U(r) \ez +  \pbu(r,\theta,z,t)$ 
where $\U(r)$ is the assumed known unperturbed streamwise velocity---the velocity field which we would observe were the pipe a smooth cylinder---and $\pbut = (\p u_r,\p u_\theta, \p u_z)$ is a small velocity perturbation due to the wall undulations.
Subscript `$\rmt$' denotes `total'. 
The Navier--Stokes and continuity equations and their boundary conditions at the wall read
\begin{subequations}
	\begin{alignat}{2}
	\left.
	\begin{aligned}
	\pp_t\pbu 
	+ ( \pbut \tcdot \nabla)\pbut + \nabla \p p  &= \Rey\inv \nabla^2\pbu\\
	\nabla \tcdot \pbu &=0 
	\end{aligned}
	\right\};&
	\quad& 0 &\leq r\leq1+\p\eta(\theta,z), \label{eq:problem:Euler}\\
	\left.
	\begin{aligned}
	\pbut \tcdot \nabla  \p\eta = \p u_r 
	&
	\\
	[\text{viscous stress closure}]&
	\end{aligned}
	\right\};&
	\quad& r &= 1+
	\p\eta (\theta,z), 
	\label{eq:problem:BC_b}
	\end{alignat}
	\label{eq:problem}%
\end{subequations}%
where we define the Reynolds number as 
$\Rey=DU_\text{avg}/\nu$ 
where $U_\text{avg}$ is average velocity, $D=2R$ diameter, and
$\nu$ the kinematic viscosity. 
In the theory we use the approximate $\Rey = RU_0/\nu$ since 
\srev{
the flow is assumed similar to normal Poiseuille flow for which $U_\text{avg}= \half U_0$.
} 
 Solutions must be smooth at $r=0$, and the basic flow is assumed to satisfy equations of motion.

Viscosity is treated in a somewhat indirect manner; it 
manifests primarily in the `zeroth--order' profile of the unperturbed current, $U(r)$, which 
\srev{
satisfies no--slip boundary conditions at $r=1$ and provides the $\order(1)$ azimuthal vorticity created by wall friction. 
}

Next, the linear--order solution is found. 
\srev{
Rather than attempt to solve an Orr--Sommerfeld-like equation in cylindrical geometry satisfying no-slip at the wavy wall (which, even if we could, would likely be too involved to be instructive) we make use of 
}
a simple model in the vein of \cite{Craik_1970_Langmuir_myidea} which captures the kinematics of how streamlines near the wall are displaced by the wavy pattern. 
Noticing that the wave--like first--order perturbation velocities are stable also in the absence of viscosity when $\p\eta$ is small, and may be assumed virtually unaffected by viscosity (this no longer holds as $\p\eta$ increases as we shall see), they approximately solve a steady inviscid and linearised form of \eqref{eq:problem}, except that an appropriate wall boundary condition must be devised. 

We assume that the boundary flow creates a displacement thickness $\sim\delta^*$ near the undulating wall and that the physical pipe wall is at $r=1+\delta^* + \p \eta (\theta,z)$. Next we impose free slip boundary conditions at a displaced boundary $r=1+\p \eta (\theta,z)$ --- see figure \ref{fig:schematic}b. Hence the shape $\p \eta$ which we specify does not quite equal the wall shape in simulations, yet while direct quantitative comparison is not possible, this model makes for a simple theory able to elucidate the nature of the Langmuir mechanism.

\srev{
Lowercase variables, which are small, are assumed to be steady and inviscid, and we expand them in powers of $a$ (formally identical to an expansion in $\ve$) according to
\be
  q(r,\theta,z) = q_1(r)\exp(\rmi m \theta + \rmi k z) + \order(\ve^2)\text{ harmonics},
\ee
(real part understood) 
}
where $q$ is any small field quantity.
\srev{
The governing linearised Euler and continuity equations \eqref{eq:problem:Euler} now read
\bs\label{eq:linEuler}
\begin{align}
  \rmi k U  u_{r,1} &= -p'_1  \label{eq:linEuler:r}\\
  \rmi k U u_{\theta,1} &= - (\rmi m/r)p_1 \\
\rmi k U u_{z,1} + U' u_{r,1}  &= -\rmi k p_1 \\
(ru_{r,1}) ' +\rmi m u_{\theta,1} + \rmi k ru_{z,1} &= 0.
\end{align}
\es
Here $m$ and $k$ real constants, the former an integer, which we will soon identify as the azimuthal and streamwise wavelengths of the imposed crossing wall waves. 
Primes denote the derivative with respect to $r$
We
}%
eliminate velocity components from 
\eqref{eq:linEuler} 
and obtain a Rayleigh--like boundary value problem for the first--order perturbation pressure $\p p_1$,
\begin{subequations}\label{eq:RayleighBC}
\begin{align}
\pnil''+\br{\frac 1r - 2\frac{\U'}{\U}}\pnil'-\br{\frac{m^2}{r^2}+k^2}\pnil = 0
\label{eq:Rayleigh} \\
	\pnil(0)=\pnil'(0)=0; ~~~~~~~~~ \pnil'(1) = [k \U(1)]^2 \eta.
	\label{eq:BC_kin}
\end{align}
\end{subequations}
\srev{
Boundary conditions for $p_1$ were found from \eqref{eq:problem:BC_b} using \eqref{eq:linEuler:r}. 
}
$\pnil(r)$ is found numerically from \eqref{eq:RayleighBC} using a standard ODE solver.

Armed with the linear order solution we proceed to the second order in $\p\eta$. 
While formalistically different due to cylindrical rather than planar geometry, the procedure is similar in outline to that of \citet{akselsen20}, hence the presentation here is comparatively briefer. 
Assume boundary undulations composed of two crossing sinusoidal waves directed symmetrically about the streamwise direction $z$:
\begin{equation}
	\p\eta = \frac{\amp}{4}\left[\rme^{\rmi( \knil z + \mnil \theta)}+\rme^{\rmi(\knil z - \mnil \theta)} 
	+ \mathrm{c.c.} \right]
	=\amp\cos(\knil z)\cos(\mnil \theta).
\label{eq:O2_modes}
\end{equation} 
We impose axial wave number $\knil>0$ and the integer azimuthal wave number $\mnil\geq 1$. 
The first-order wave modes involved each have amplitudes $\amp/4$ and the four wave vectors $(\pm \knil,\pm \mnil)$ (signs varied individually). Second order harmonics, in turn, obtain wave vectors which are sums of pairs of these, thus being of four different types with wave vectors $\pm 2(\knil,\mnil)$, $(0,0)$, $(\pm 2\knil,0)$ and $(0,\pm 2\mnil)$. 
The three first types remain of order $a^2$ and can be neglected, whereas we retain the last type of harmonic, which 
\srev{
turns out to be
}
resonant with a wave vector modulus $2\mnil$,
and grows linearly with time as $a^2 t$ until further development is checked by viscous damping 
\srev{
(the resonant, linearly growing solution is given in appendix \ref{appendix}; an extensive discussion for the planar sibling system, see \citet{akselsen20}).  
}
The resonance will manifest in the formation of $\mu=2\mnil$ pairs of streamwise vortices as sketched in Figure~\ref{fig:schematic}a. 
All second order fields henceforth are understood to be of form $\uq_2(r,\theta,z,t) = \uq(r,t)\exp(\rmi \mu \theta) + \mathrm{c.c.}$ with $\uq\in\{\uu_r,\uu_\theta,\uu_z,\up\}$; note that these are independent of $z$,
\srev{
and hence constitute secondary motion in the $(r,\theta)$ plane. 
The second--order Navier--Stokes and continuity equations then read
\bs
\begin{align}
  \D \uu_r + \frac{2\rmi \mu}{r^2 \Rey}\uu_\theta + \pp_r \up =& -[(\bu_1\cdot\nabla)\bu_1]_r, \\
  \D \uu_\theta - \frac{2\rmi \mu}{r^2 \Rey}u_r + \frac{\rmi \mu}{r} p =& -[(\bu_1\cdot\nabla)\bu_1]_\theta, \\
  \D \uu_z + \U'(r) u_r - \frac{1}{r^2 \Rey}u_z =& -[(\bu_1\cdot\nabla)\bu_1]_z, \\
  (r\uu_r)' + \rmi \mu u_\theta =& 0.
\end{align}
\es
Here, $\D  =  \pp_t  -\Rey^{-1} [  \pp_r^2 + r^{-1} \pp_r - r^{-2}(1+\mtwo^2)]$.

We find it most convenient now to work with the radial velocity component. 
}
Upon eliminating the second order 
axial and azimuthal velocities 
and pressure 
one retrieves an inhomogeneous Orr--Sommerfeld-type equation
\begin{gather}
 \frac{1}{r^2}\pp_r\wigbrac{r \D \sqbrac{\pp_r\br{r \uu_r}}}
- \frac{\mtwo^2}{r^2}\br{\D - \frac{4}{r^2 \Rey}}\uu_r = \mc R(r); 
\label{eq:Rayleigh_O2}
\\
\mc R(r) = 8\br{\frac{\mnil}{r\knil \U 
}}^2\frac{\U'}{\U}\sqbrac{\br{\knil^2-\frac{\mnil^2}{r^2}}\pnil^2 + (\pnil')^2}
\label{eq:R}
\end{gather}
for the radial second order velocity $\uu_r(r,t)$.
\srev{
Note that $\mc R$ and $\uu_r$ are proportional to $a^2$. 
}

Equation \eqref{eq:Rayleigh_O2} permits fairly simple analytical solutions in the two opposite cases of transient inviscid flow ($\Rey\inv= 0$) and stationary viscous flow ($\pp_t\,\cdot\rightarrow0$) representing onset and ultimate stages of vortex development, respectively. We consider here only the latter which will inform the steady--state reached in simulations. 
For completeness, the solution for initial growth rate is presented in 
\srev{
an appendix.
}

Assuming a steady state with finite $\Rey$, \eqref{eq:Rayleigh_O2} has solution 
\begin{align}
\uu_r(r) &= 
\frac{r^3 \Rey}{8 \mtwo}\sum_{s=\pm1}
\bigg\{
\sum_{\s=\pm1} \frac{1}{\s+s \mtwo}
\int_1^r \!
\dd \rho\,
\br{\frac \rho r}^{s+ \s \mtwo+3} \mc R (\rho)
\nonumber\\&
-\int_1^0 \!
\dd \rho\,\sqbrac{\frac{1}{1+s\mtwo} +  \frac{s r^{2\mtwo}}{\rho^{1+s}} \br{1-\rho^2-
\frac{1-\rho^2+s(1+\rho^2)}{2(1-s\mtwo)}} } \br{\frac{\rho}{r}}^{s+\mtwo+3} \mc R (\rho)
\bigg\}
\label{eq:urO2_sol_viscous}
\end{align}
where no-slip boundary conditions at the wall are imposed. The streamwise velocity is
\begin{equation}
\uu_z(r) = 
\frac{\Rey}{2\mu} \sum_{s=\pm1} s \br{
\int_1^0\!
\dd\rho\,\frac{\rho^{\mtwo}}{r^{ s\mtwo}} -
\int_1^r\!
\dd\rho\, \frac{\rho^{ s\mtwo}}{r^{s\mtwo}} 
}\rho \,\U'(\rho) \uu_r(\rho)  .
\label{eq:u_z_viscous}
\end{equation}
Thus  the radial and streamwise velocity perturbations scale as $\Rey$ and $\Rey^2$, respectively.
Assuming $U'(r)<0$, $\uu_r$ and $\uu_z$ are of a sign, so the secondary motion accelerates the mean flow in areas where the circulation jets towards the wall, and \emph{vice versa}.

The second-order vortical motion being independent of $z$ we introduce a stream function 
$\psi$ whose contours are streamlines. By definition $\uu_r = r^{-1} \pp_\theta\p\psi$ and $\uu_\theta = -\pp_r\p\psi$.
In terms of the stream function amplitude  $\upsi(r) = 2r\uu_r/\mu$ we find
\begin{align}
\p\psi(\theta,r) &= \upsi(r) \sin(\mtwo \theta),
&\p u_z(\theta,r) &= 2\, \uu_z(r)  \cos(\mtwo \theta)
\end{align}
from which $\p u_\theta$ can be inferred if required.


\section{Theoretical predictions}
\label{sec:results}

\begin{figure}%
\includegraphics[width=\columnwidth]{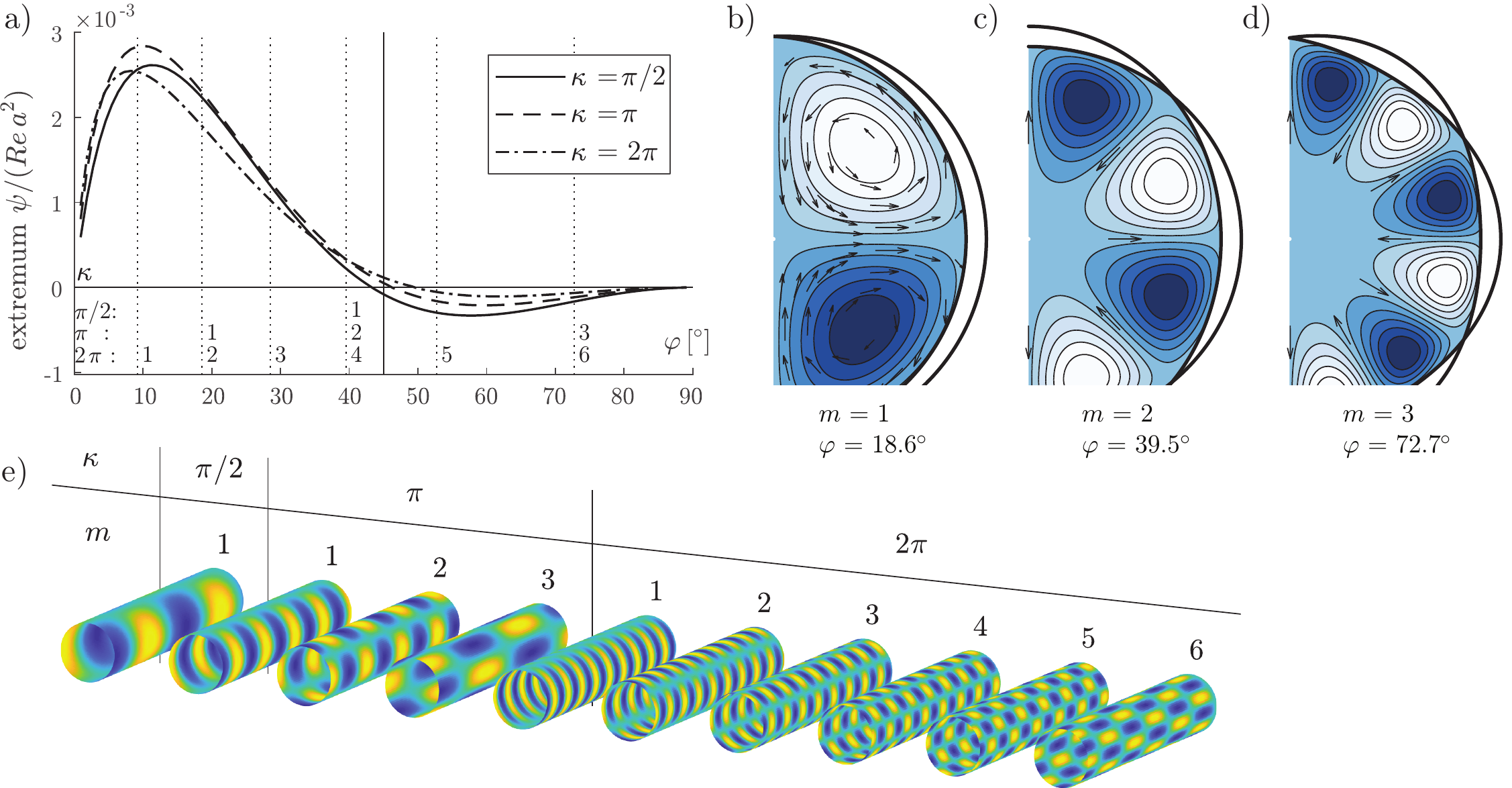}
\caption{ Theoretical predictions. (a) Circulation intensity for fixed $\kappa=(\knil^2+\mnil^2)^{1/2}$ as function of phase angle $\thetanil$; $\delta^*=0.05$. 
\srev{
Only design configurations for which $m$ is an integer are realisable; these are marked with dashed vertical lines 
}
where $\mnil$ values are marked as integers. 
(b-d): Streamlines in the cross-flow plane 
\srev{
for $m\in \{1,2,3\}$ and $\kappa=\pi$, which are contours of  $\p\psi$. 
}
Velocity field vectors are shown for $\mnil=1$ whereas arrows in (c,d) merely indicate flow direction. Circulation intensity may be inferred from (a)  
considering $\kappa=\pi$. 
\srev{
Pipe cross section outlines are shown at the crests/troughs of the wavy pattern ($z=\lambda/4$ and $3\lambda/4$ with $\lambda=2\pi/k$). Colours illustrates the value of $\psi$ with light (dark) being positive (negative).
}
(e): Pipe design configurations corresponding to the dashed vertical lines in panels a and b; $45^\circ$ is marked with a solid vertical line. 
}%
\label{fig:theory}%
\end{figure}

While the theory in section \ref{sec:theory} is simplistic and captures only one of the causes of secondary flow, its predictions are instructive and will inform our DNS study below. We consider only $m_1\leq 3$ below; higher values create more and smaller vortices closer to the wall but there is no indication of further change of behaviour.

Assume a laminar bulk flow profile of Poiseuille type, 
\begin{equation}
\U(r) =  1 - r^2/(1+\delta^*)^2
\label{eq:parabolic_profile}
\end{equation}
stretched a displacement length $\delta^*$ beyond the pipe radius as sketched in figure~\ref{fig:schematic}b.
Henceforth we use the term \emph{crestline} to denote a curve following the wall at constant polar angle $\theta = n\pi/m_1, n=0,...,2m_1-1$, running over the maxima of crests and troughs, and \emph{saddlepoint line} for the nearly straight line following the wall midway between these. Streamlines close to crestlines have the largest undulations in wall--attached flow.

A key parameter is the angle $\thetanil = \arctan(\mnil/\knil)$ between the streamwise and azimuthal wavenumbers of the wall undulation which we refer to as the crossing angle. We let $0\leq \varphi \leq 90^\circ$. 
We shall refer to geometries $\thetanil < 45^\circ$, $=45^\circ$ and $>45^\circ$ as contracted, regular and protracted egg carton patterns, respectively. The theoretical dependence of circulation strength on $\thetanil$ is investigated in Figure~\ref{fig:theory} where the wave vector modulus $\kappa =(\knil^2+\mnil^2)^{1/2}$ is kept constant at three different values while $\thetanil$  changes. 

Figure~\ref{fig:theory}a shows the steady--state circulation strength, represented by the extremum of $\psi/Re\, a^2$
along a ray at $\theta=\pi/4\mnil$.
The integer $\mnil$ can only take values $1,2,...,\mathrm{floor}(\kappa)$,
shown with vertical lines labelled with corresponding azimuthal wavenumber $\mnil$. Corresponding pipe patterns are shown in Figure \ref{fig:theory}e.
The volume flow rate through a vortex cross-section is proportional to  $\mathrm{max}|\psi|$, and the sign of $\psi$ shows the rotational direction:  relative to the pipe wall $\psi>0$ indicates flow towards crestlines and away from saddle--point lines, and \emph{vice versa}. 

Several observations are made. The circulation intensity is 
\srev{relatively} 
insensitive to wavenumber amplitude $\kappa$ but highly sensitive to $\thetanil$. 
The Langmuir driving mechanism is very weak near $\thetanil = 45^\circ$, the only angle previously investigated for pipe flow to our knowledge, and $\p\psi$ changes sign near this angle. 
(We note in passing that the secondary flow observed in turbulent pipe flow at $\thetanil=45^\circ$ by \citet{chan18}, corresponded to negative $\psi$. )
Moreover, the intensity of the `reversed' Langmuir rotation at $\thetanil>45^\circ$ is considerably weaker than that predicted for smaller angles $\thetanil\lesssim 30^\circ$.

Figure~\ref{fig:theory}b--d shows streamlines $\psi=\const$ of the flow averaged over an axial wall wavelength, 
for the three possible angles when $\kappa=\pi$. 
Notice again the reversal of rotation direction for $m=3$ where the pattern is protracted.


\section{Simulations}
\label{sec:sim_results}

We proceed now to study the real flow in the wavy pipe geometry using DNS,  
focussing on the effects of wave crossing angle $\thetanil$, Reynolds number and topography amplitude. Further plots and figures  for all simulation cases may be found in Supplementary Materials.
Velocities are in units of the mean centreline velocity for each case.

\begin{figure}%
\centering
\includegraphics[width=\textwidth]{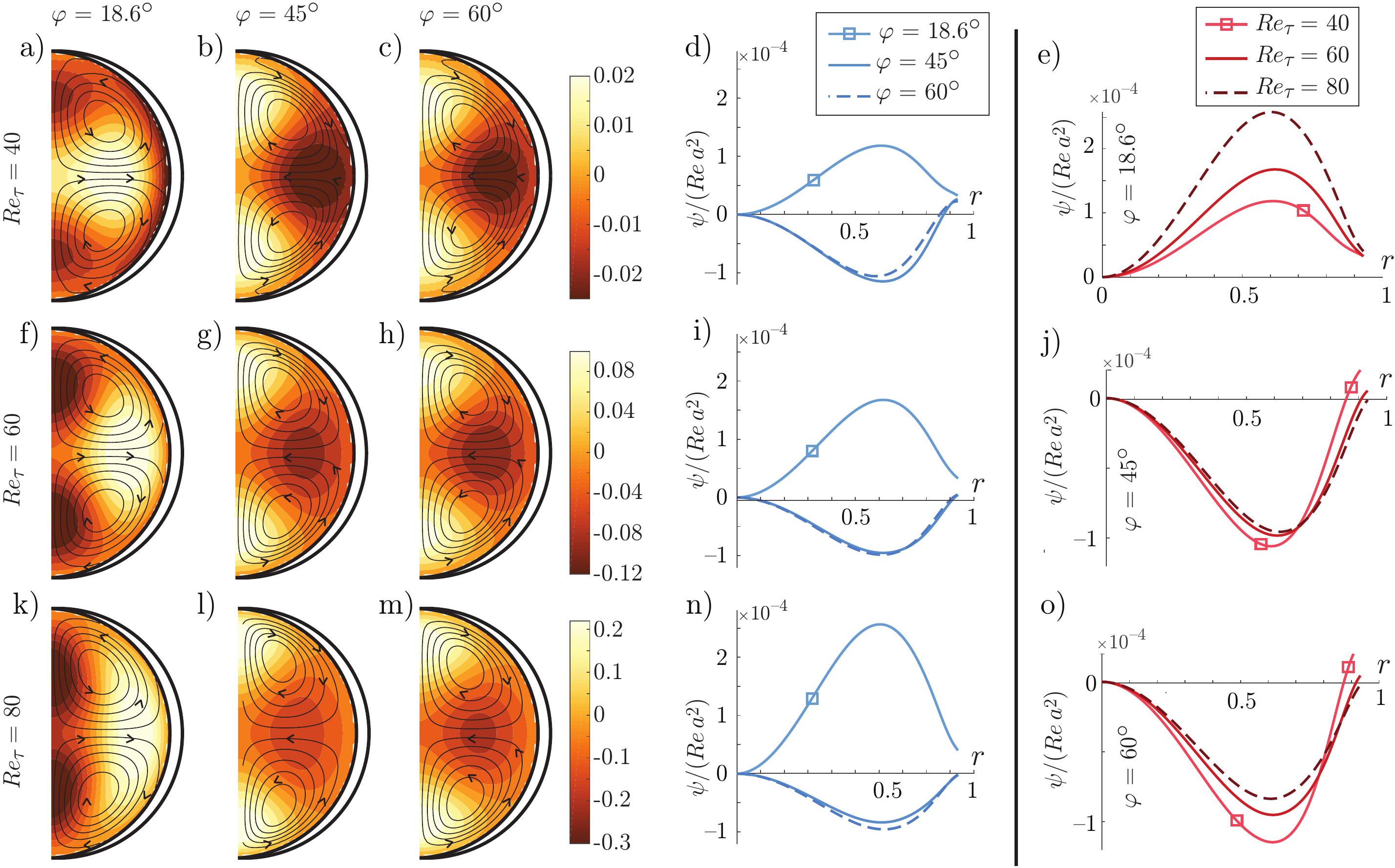}
\caption{
Simulation results, $a=0.05,\kappa=\pi$ and $m=1$. 
(a-c,f-h,k-m): Black curves are contours of $\acpsi/(\Rey\, a^2)$ between $\pm 2.5\cdot 10^{-4}$, arrows indicate flow direction; colour contours show deviation of $\oluzt$ from Poiseuille flow. (d,e,i,j,n,o): plots of $\acpsi/(\Rey\, a^2)$ along the ray $\theta=\pi/4$. A common legend applies to all panels where $\varphi$ varies at constant $\Rey$ (d,i,n), and \emph{vice versa} (e,j,o).
}%
\label{fig:psi}%
\end{figure}

The numerical simulations were conducted using NEK5000, a high-fidelity spectral element code \citep{nek5000-web-page}. Each computational domain contains $1280$ macro lements with $10$ macro elements in the streamwise direction. The nodes inside of the element are distributed using the Gauss–-Lobatto-–Legendre (GLL) points and a polynomial order of $7$ is used, resulting in approximately $655\,360$ grid points in total. The grid points on the no-slip, impermeable wall of the pipe conforms to the roughness topography, the domain length equal to one roughness period and the ends of the pipe are periodic. The 3rd order time-stepping scheme and the $P_N-P_{N-2}$ method introduced by \citet{maday1989spectral} was used for the simulations.  
A constant pressure gradient is used to drive the flow and the simulations were run with a constant timestep ranging 
$\mr{d} t^+ = t \, U_\tau^2/\nu = 10^{-4}$ to $2\times10^{-4}$ 
($U_\tau = \sqrt{\tau_w/\rho}$ is the friction velocity, $\tau_w$ the 
mean wall shear stress) to ensure that the Courant number is less than 1. The simulations were initialised with a laminar smooth--wall flow and were run for a duration of at least $t^+ = 1600$ where the flow has converged to a steady state.
\srev{
The grid points on the no--slip, impermeable wall of the pipe conforms to the roughness topography and the ends of the pipe are  periodic. The length of the domain is equal to one roughness period. A domain length study conducted for $\varphi = 18.6$ with a = 0.05 at $Re_\tau = 80$ and no changes to the steady-state flow was observed when the length of the pipe was increased by $6$ and $10$ times. 

}

One primary observation we make through a broad parameter study in this section is that a competition occurs between two effects, both of which driving secondary motion, directed oppositely. One is a dynamic effect due to increased wall shear stress where the roughness is increased near crestlines, the other the kinematic Langmuir circulation effect, CL1. The former causes secondary flow in the negative sense as defined, the latter drives positive-sense rotations for $\varphi\lesssim 30^\circ$ where it is strongest according to theory.

\srev{
It is highly useful for our further analysis to introduce streamwise--averaged quantities. Noting that our flow is steady and periodic with streamwise period (or wavelength) $\lambda = 2\pi/k_1$ we define the averaging operator
\be\label{average}
  \ol{(\cdots)} = \frac1{\lambda}\int_0^{\lambda}(\cdots)\rmd z
\ee

}

Based on the principle of volume flux, a measure of circulation strength in the simulated flows is found as the approximate stream function amplitude $\psi$ along a radial line of constant polar angle $\theta=\theta_0$ running through, or nearly through, the centre of a vortex. We choose $\theta_0=\pi/4m_1$ which approximately bisects the `first' vortex. We define
\begin{equation}
	\acpsi(r;\theta_0) = \int_{0}^r \!\dd \rho\, \overline{\p u_\theta}(\theta_0,\rho).
	\label{eq:tilde_psi}
\end{equation}

\subsection{Parameter studies}

Dependence on crossing angle $\varphi$ and $\Rey$ is studied in Figure \ref{fig:psi}; rows have constant $\Rey$, columns constant $\varphi$. All graphs are of $\acpsi/\Rey\, a^2$. Note that in all plots of quantities averaged over a streamwise wave period, linear effects of wall undulations vanish and only contributions from (even) higher-orders remain.

We investigate three different crossing angles, $\varphi=18.6^\circ, 45^\circ$ and $60^\circ$. According to theory, Langmuir motion should be strongest and positive for the first angle, and much weaker for the two latter; see figure \ref{fig:theory}b. Indeed, the most striking feature in figure \ref{fig:psi} is arguably that the smallest angle shows positive circulation (first column: a,f,k), the other two negative (second and third columns: b,c,g,h,l,m). 
However, unlike in the theoretical graph of the Langmuir effect alone, figure \ref{fig:theory}, the oppositely directed circulation at $45^\circ$ and $60^\circ$ is not weak, but of comparable magnitude as for $18.6^\circ$, evidence of another mechanism at play. 
We propose that there is a dynamic, viscosity--driven forcing of negative circulation present due to the asimuthally varying roughness producing alternating regions of higher and lower momentum as observed by \citet{chan18},  
which depends only weakly on $\varphi$. The competing Langmuir effect is significant only for the smallest angle. Indeed, in all simulations, the flows at $45^\circ$ and $60^\circ$ are highly similar, whereas $18.6^\circ$ flow is strikingly different (see also supplementary material). 
%


\begin{figure}%
\centering
\includegraphics[width=\columnwidth]{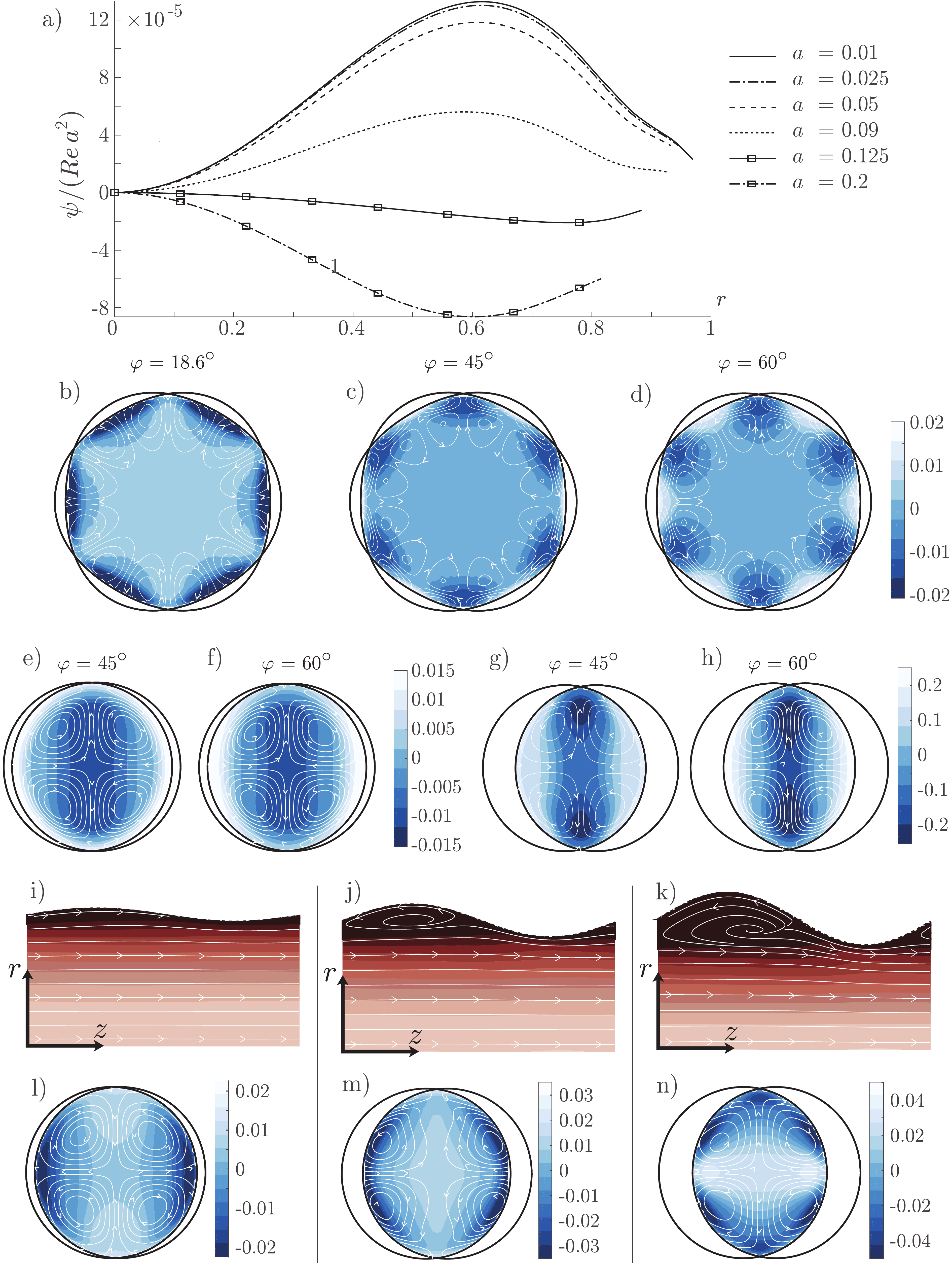}%
\caption{Simulation results. (a) Scaled circulation strength along $\theta=\pi/4m_1$ for $\thetanil = 18.6^\circ, m_1=1$ and increasing $a$; graphs from top to bottom: $a=0.01, 0.025, 0.05, 0.09,0.125,0.2$. Panels (b-h, l-n): streamwise--averaged flow (streamlines) and pressure (colour contours, average pressure subtracted). Panels (b-d): $m_1=3$, all other panels $m_1=1$. Crossing angle $\thetanil=18.6^\circ$ except as indicated. 
Amplitudes are $a=0.05$ (b-f,i,l), $a=0.125$ (j,m), $a=0.2$ (g,h,k,n). (i-k): streamlines in a streamwise section of the pipe through crests/troughs at $\theta=\pi/2$, lighter (darker) colour indicates higher (lower) average absolute velocity. 
}%
\label{fig:ampl}%
\label{fig:pressure}%
\end{figure}

\subsubsection{Sensitivity to Reynolds numbers and crossing angle $\thetanil$. } 

We define $\Retau=U_\tau R/\nu$ and $\Rey=U_\mathrm{avg}D/\nu$ where $U_\mathrm{avg}$ is total flow rate divided by $\pi R^2$. For Poiseuille flow, $\Rey= \half Re_\tau^2$. 

Figure~\ref{fig:psi} shows simulation results for $a=0.05$ and $m=1$, varying $\varphi$ along rows and $\Rey$ along columns. Three different topography angles $\thetanil = 18.6^\circ$, $45.0^\circ$ and $60.0^\circ$ --- contracted, regular and protracted eggcarton, respectively --- are shown, 
and three different $\Retau=40,60,80$. All panels show values of $\tilde\psi(r,\pi/4m_1)/(\Rey\, a^2)$ either as contours or graphs. The highest Reynolds achieved in reported simulations is $2751$. A simulation at $\Retau=100$ became turbulent (not included since the grid used herein is too coarse to properly capture turbulent flow). Our simulations are not sufficient to draw confident conclusions about stability in each case, which remains a question for the future.

Studying the rightmost column of Figure~\ref{fig:psi}, we observe that the expected scaling $\psi\propto \Rey$ is reasonably well satisfied throughout the laminar regime for regular and protracted eggcarton, whereas for the contracted eggcarton the scaling is far more imperfect. 
\srev{
In fact, for $\thetanil = 18.6^\circ$, $\psi$ increases faster than linearly with $\Rey$, a curious observation we discuss further in section \ref{sec:drag}.
} 
The departure from the scaling predicted by inviscid theory can be traced back to a greater deviation between the theoretical, inviscid first-order velocity field and that from simulations, an indication that viscous effects in the boundary layer influence the results considerably in a non-trivial way, more strongly for the contracted pattern. 
\srev{
A partial explanation is that, for one and the same $\kappa$, smaller $\varphi$ corresponds to higher steepness $\ve=\kappa a (1-\sin^2\varphi)^{1/2}$, and higher--order non--linear effects manifest more easily. We subject this curious observation to closer scrutiny in section \ref{sec:drag}. 
}

It is instructive to regard the pressure field across the pipe section when averaged along a streamwise wavelength so that linear order perturbations vanish leaving a mean pressure deviation able to drive steady secondary motion. Compare the pressure fields in Figure~\ref{fig:ampl}e,f and l wherein $\varphi=45^\circ, 60^\circ$ and $18.6^\circ$, respectively, for $a=0.05$. The flow and pressure perturbations for the two former are similar: high pressure regions above crestlines push the flow away from the wall there, driving vortices in the negative sense. This is as might intuitively be expected since the flow suffers higher friction here than along the straighter saddlepoint lines. The pressure field for $18.6^\circ$ on the other hand shows the opposite: low-pressure regions above crestlines attract the secondary flow setting up positive--sense vortices.

Our suggested interpretation is as we began to argue above. The dynamic friction mechanism evident in Figure~\ref{fig:ampl}e,f will be present for all three values of $\varphi$ in roughly equal measure; the strong similarity between Figures~\ref{fig:ampl}b and c indicates that it varies little with $\varphi$ so long as the flow does not separate. On the other hand, the Langmuir mechanism is far stronger for $\varphi=18.6^\circ$ than for the two higher values (see Figure~\ref{fig:theory}a), and therefore `wins' the competition there. 
We note with interest, and for future reference, that the swirling changes sign close to the pipe wall for $\varphi=45^\circ, 60^\circ$ and $\Retau=40,60$.

\subsubsection{Sensitivity to amplitude. } 
\label{sec:ampl}

\newcommand{\tpsi}{\tilde\psi} 

Interestingly, when increasing the amplitude $a$, circulation reversal is observed for $\thetanil=18.6^\circ$. We again propose an explanation in terms of the two competing mechanisms for secondary flow. In Figure~\ref{fig:ampl}a we plot the scaled circulation strength $\tpsi/\Re \,a^2$ in the protracted eggshell geometry for increasing amplitudes up to $a=0.2$. The predicted $\sim a^2$ scaling is accurate for moderate amplitudes $a \leq 0.05$, but beyond this point a dramatic reduction occurs, and as $a\gtrsim 0.1$ the direction of rotation reverses with $|\tpsi|/a^2$ eventually reaching comparable values. 

We find the reason to be the onset of flow separation affecting the two mechanisms differently. Roughly, the Langmuir swirling is driven by the kinematic sinusoidal deflection of streamlines; once the flow separates in the troughs, streamlines no longer follow the wall's shape (see Figure~\ref{fig:ampl}h-j) and a further increase in $a$ does not further increase the `effective amplitude' of the streamline undulations. For $\thetanil=45^\circ$ and $60^\circ$ the wall undulations are less steep in the streamwise direction and the flow does not separate, retaining the $\sim a^2$ scaling.

\newcommand{\Uav}{U_\text{avg}}
\newcommand{\rmP}{\mathrm{P}}
\srev{

\subsubsection{Circulaton strength vs increased drag}
\label{sec:drag}

\begin{figure}
  \includegraphics[width=\textwidth]{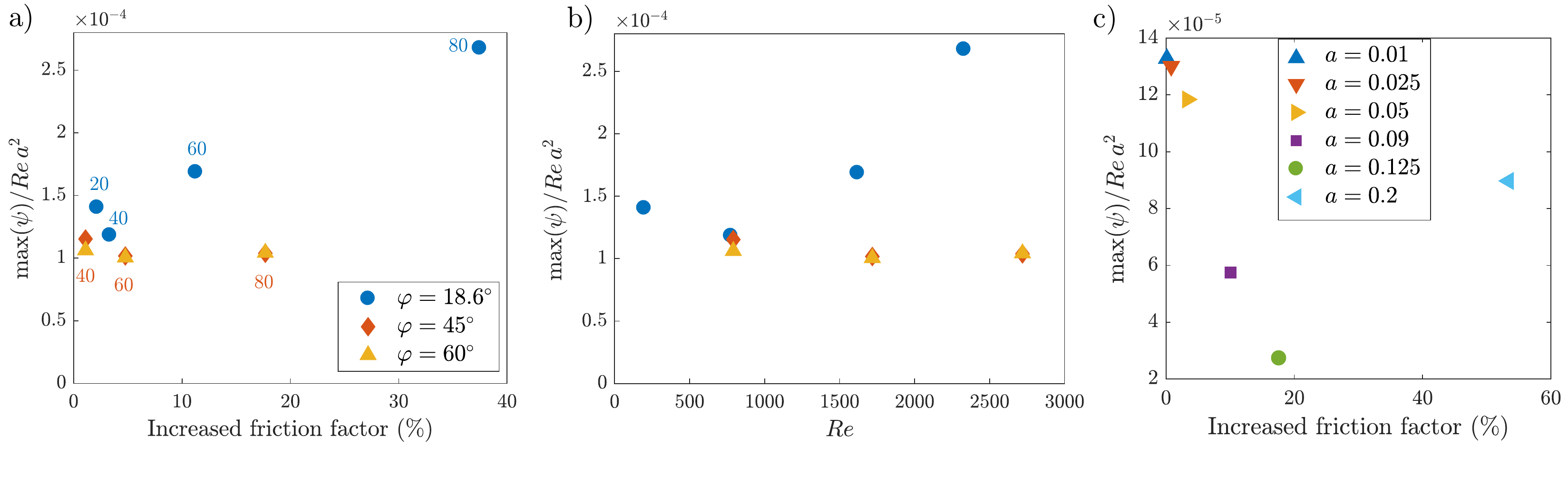}
  \caption{Maximum circulation strength plotted against increase in friction factor (i.e., increased head loss) in percent. a) Comparison of three different crossing angles for $m=1, a=0.05$ --- corresponding values of $\Rey_\tau$ are indicated for each marker (common for overlapping markers); b) same cases as in panel a, but with Reynolds number as abscissa; c) increasing amplitude for $\Rey_\tau=40$ and $m=1$.}
  \label{fig:drag}
\end{figure}

It is of interest to compare the strength of circulation to the increased pressure loss from making the wall surface wavy. Let 
$\Uav$ be the streamwise velocity averaged over the cross-section. Using the definition of the Darcy friction factor $f=(2gD/\Uav^2)h_L$, $h_L$ being the head loss per streamwise wavelength related to $\tau_w$ by $h_L=2\tau/\rho g R$ (dimensional units), gives 
\be
   f = \frac{32 \Rey_\tau^2}{\Rey^2}
\ee
having used $\Rey_\tau=(R/\nu)\sqrt{\tau/\rho}$ and $\Rey=\Uav D/\nu$. 
We will compare with Poiseuille flow with the same Reynolds number,
\be
  U_\rmP(r) = 2U_\text{avg}(1-r^2)
\ee
for which it is readily shown that $\Rey^2_\rmP=\half \Rey_\tau^2$ and $f_\rmP=64/\Rey$. The relative increase in friction coefficient is thus $(\half \Rey_\tau^2-\Rey)/\Rey$ which we plot in \% as abscissa in figure \ref{fig:drag}. 

A particularly striking observation can be made from figure \ref{fig:drag}a, where three different crossing angles $\varphi$ are compared for $a=0.05$ and $m=1$. For each angle, each marker corresponds to a different $\Rey_\tau=40,60,80$ increasing from left to right; for $\varphi=18.6^\circ$ also $\Rey_\tau=20$ is included. The points are too few to fully determine scaling, yet it appears that whereas for the two larger angles where CL1 is weak the scaled circulation strength $\max(\psi)/\Rey$ saturates to a constant value, indicating that absolute circulation strength increases as $\sim \Re$ throughout the laminar regime. For the smallest angle with strong Langmuir forcing, however, circulation strength increases faster than $\sim \Rey$, something which becomes even clearer when plotted against $\Rey$ as in figure \ref{fig:drag}b (The faster-than linear scaling was already observed in figure \ref{fig:psi}e). 

The non-monotonous dependence of circulation strength on amplitude previously discussed in section \ref{sec:ampl} is illustrated once more in the scatterplot of figure \ref{fig:drag}c.

\subsubsection{High-- and low--momentum channels}
\label{sec:hmp}

High-momentum paths (HMP) and low momentum paths (LMP) are conspicuous in figure \ref{fig:psi} where colour contours of 
\be
  \oluz(r,\theta)=\oluzt(r,\theta)-U_\rmP(r)
\ee 
are shown. Here $U_\rmP(r) = 2U_\text{avg}(1-r^2)$ is a Poiseuille flow of the same volume flux.
Both for $\varphi=45^\circ$ and $60^\circ$ the intuitively expected behaviour is seen: lower (higher) momentum resides over crestlines (saddlepoint lines) where the roughness is highest (lowest). At $18.6^\circ$ the picture is opposite, yet a telling observation is made in figure \ref{fig:psi}a: in a thin layer over the crestline wall a strong velocity deficit from increased friction \emph{is} in fact present, but is soon overtaken by CL1 away from the wall (in panels f,k the layer is so thin as to fall outside the plotted area). 
This is another indication that the two effects are simultanesouly present and competing. In all cases we note that the rotating motion is directed away from the wall where there is a low--momentum path, and \emph{vice versa}.

In studies of turbulence over spanwise varying roughness of different kinds, secondary motion has also consistently been directed away from the wall over LMPs and \emph{vice versa} irrespective of the kind of roughness \citep[e.g.][]{anderson15, willingham14,hwang18,vanderwel15,chan18, chung18}. 
\citet{colombini95} show that the situation is more subtle when a free surface is present, and \citet{stroh20} found a richer pattern of secondary motion when spanwise roughness variatons do not create a clear distinction between the two. 
While we should be careful to infer too much from turbulent mean flow to the present laminar case, it is consistent with our observations. [We bear in mind the related, but not identical, rule of thumb due to \citet{hinze67} that secondary flow is directed out of (into) areas with net production (dissipation) of turbulent kinetic energy, by which \citet{hwang18} explain the apparent inconsistency in sense of rotation of secondary flows for different types of roughness, compared to, e.g., \cite{wang06}.]

The direction of swirling for our laminar case is indicated by the streamwise averaged equation of motion.
Into the $z$--component of the Navier--Stokes equation \eqref{eq:problem:Euler} we insert $\oluzt = U_\rmP(r)+\oluz$. We use rectangular co--ordinates, but  notice that $(\olu \pp_x+\olv \pp_y)U_\rmP = \olur U_\rmP'(r) = -4\Uav r \olur$. 
Ordering in powers of $a$, applying streamwise averaging \eqref{average} and neglecting terms of $\mathcal{O}(a^2)$ yields
\be
  -2\Rey\,r\olur = \nabla^2\oluz
\ee
Near a high--momentum path where $\oluz$ has a maximum, $\nabla^2\oluz<0$ and hence $\olur>0$, and for a low--momentum path the opposite is true, thus flow is towards the wall near a high--momentum path and \emph{vice versa}. We note from the presence of $\Rey$ that this $\order(a)$ mechanism depends on the presence of viscosity.

Already we see that it is not surprising that flow is directed upwards from crestlines and down towards saddlepoint lines when Langmuir driving is weak (e.g.\ for $\varphi=45^\circ$ and $60^\circ$): fluid paths going over crests and troughs suffer higher friction than the nearly straight saddlepoint streamlines, giving rise to a low momentum channel pushing the flow towards the centre.

}

\srev{

\section{Analogy of secondary flow in turbulence}
\label{sec:Prandtl}

\citet{prandtl52} famously divided secondary flow in turbulence into two categories, now referred to as Prandtl's secondary flow of the first and second kind, respectively. 
The former stems from inviscid skewing of the mean flow, typically from the flow being guided by a curved surface; the second kind is driven by the inhomogeneity of Reynolds stresses.

It is commonly stated that Prandtl's secondary flow of the second part has no counterpart in laminar flow \cite[e.g.][p.~54]{bradshaw87}. We argue that this might be open to discussion since 
we shall see that in streamwise--periodic flow a close analogy is achieved when Reynolds averaging replaced by streamwise averaging, equation \eqref{average}. 

 The velocity and vorticity fields may be divided into mean and an oscillating parts,
\be
  \bu = \olbu + \tbu; ~~ ~ \bom =\olbom + \tbom; 
\ee
with $\bu = (u_r, \ut, u_z)$ or $(u,v,w)$, and $\bom = \nabla\times\bu= (\omr, \omt, \omz)$ or $(\omx,\omy,\omz)$, with accents as appropriate. 

Let $\vartheta$ denote any field quantity henceforth. Note the relations
\bs
\begin{align}
  \ol{\tilde{\vartheta}} =& 0; \\
 \ol{\pz \vartheta} =& 0; \label{meandz}\\
  \ol{\pp_i \vartheta}=&\pp_i\ol{\varphi},
\end{align}
\es
where $i\in \{ x,y\}$ or $\{r,\theta\}$.

For simplicity we work first in rectangular co--ordinates; the direction of mean flow remains $z$. 
Consider the streamwise component of the vorticity equation. Exactly following the procedure of, e.g., \citet{anderson15} but for the definition of the averaging operator, one finds (with $\nabla_\perp^2 = \px^2 + \py^2$)
\be\label{vorz_1}
  (\olu\px + \olv \py)\olomz = \olomx\px \olw+ \olomy\py \olw+ (\py^2 - \px^2)R_{xy}+ \px\py(R_{xx} - R_{yy})+\nu\nabla_\perp^2\olomz
\ee
where we define the \emph{undulation stress}
\be
  R_{xx} = \ol{\tu\tu};~~~ R_{yy} = \ol{\tv\tv};~~~ R_{xy} = \ol{\tu\tv}.
\ee
Replacing streamwise averaging with Reynolds averaging, equation \eqref{vorz_1} is a classic one \citep{bradshaw87}. 
The undulation stresses are analogous to what in turbulence is often dubbed \emph{dispersive} stress \citep{raupach82} arising from spatial correlation of time--averaged quantities; we eschew this term to avoid any confusion with dispersion of surface waves, featuring in the literature on Langmuir circulations. 

The first two terms on the right--hand side of \eqref{vorz_1} would in turbulent flow correspond to Prandtl's first kind of secondary flow. These add to zero in streamwise--periodic flow which is obvious once we note that $
  \olomx = \py \olw$ and
$\olomy = -\px \olw$.

\newcommand{\omdiss}{\nu\nabla_\perp^2\olomz}
\newcommand{\Sdef}{S_\text{def}}
\newcommand{\Sadv}{S_\text{adv}}
\newcommand{\Sprod}{S_\text{prod}}

We are left with the terms involving the undulation stresses, which may be written in the following two forms
\bs\label{vorz}
\begin{align}
  (\olu\px + \olv \py)\olomz =& S_\text{norm} +S_\text{shear}\label{vorza}+\omdiss\\
  =&  S_\text{def} + S_\text{adv}\label{vorzb}+\omdiss.
\end{align}
\es
with 
\bs
\begin{align}
   S_\text{norm} =& \px\py(R_{xx} - R_{yy}); ~~~
   S_\text{shear} = (\py^2 - \px^2)R_{xy};\\
   \Sdef =&  \ol{(\tbom\cdot\nabla)\tw} = \half S_\text{norm} + S_\text{shear} + \px\ol{\tu\px\tv}-\py\ol{\tv\py\tu}\notag \\
      =& \half S_\text{norm} +\py\ol{\tu\py \tv}-\px\ol{\tv\px \tu};\\
   \Sadv =&- \ol{(\tbu\cdot\nabla)\tomz}= \half S_\text{norm} - \px\ol{\tu\px\tv}+\py\ol{\tv\py\tu}.
\end{align}
\es
We let the total mean vorticity production be
\be
  \Sprod = S_\text{norm} + S_\text{shear} = \Sdef + \Sadv.
\ee

The form \eqref{vorza} is the standard in the turbulence literature, and has the advantage of highlighting the asymmetry of $R_{ij}$ under $x\leftrightarrow y$ as the explicit cause of streamwise vortices, due to normal and shear stresses, respectively. While a natural choice in wall and channel type geometries, in our present case we find a physical interpretation of the individual terms less obvious. Going to cylindrical co-ordinates mixes the roles of normal and shear stresses: by expressing $u,v$ in terms of $u_r, v_r$ and $\theta$ 
one finds, 
\be
  R_{xx}-R_{yy} = 2 R_{r\theta}\sin 2\theta; ~~ R_{xy} = -2R_{r\theta}\cos 2\theta+\half(R_{rr}-R_{\theta\theta})\sin 2\theta
\ee
with $R_{rr}=\ol{\tur^2}, R_{\theta\theta}=\ol{\tut^2}$ and $R_{r\theta}=\ol{\tur\tut}$. 
Some light might be shed from recasting the full analysis in cylindrical coordinates, but we choose instead to analyse vorticity transport in terms of \eqref{vorzb}, the form favoured by \citet{nikitin19}. 

The form \eqref{vorzb} is advantageous in that the two production terms $S_\text{def}$ and $S_\text{shear}$ are independent of choice of co-ordinate system. Physically they correpond, respectively, to production of streamwise--average vorticity by periodic deformation and advection of vorticity, respectively.

\begin{figure}
  \includegraphics[width=\textwidth]{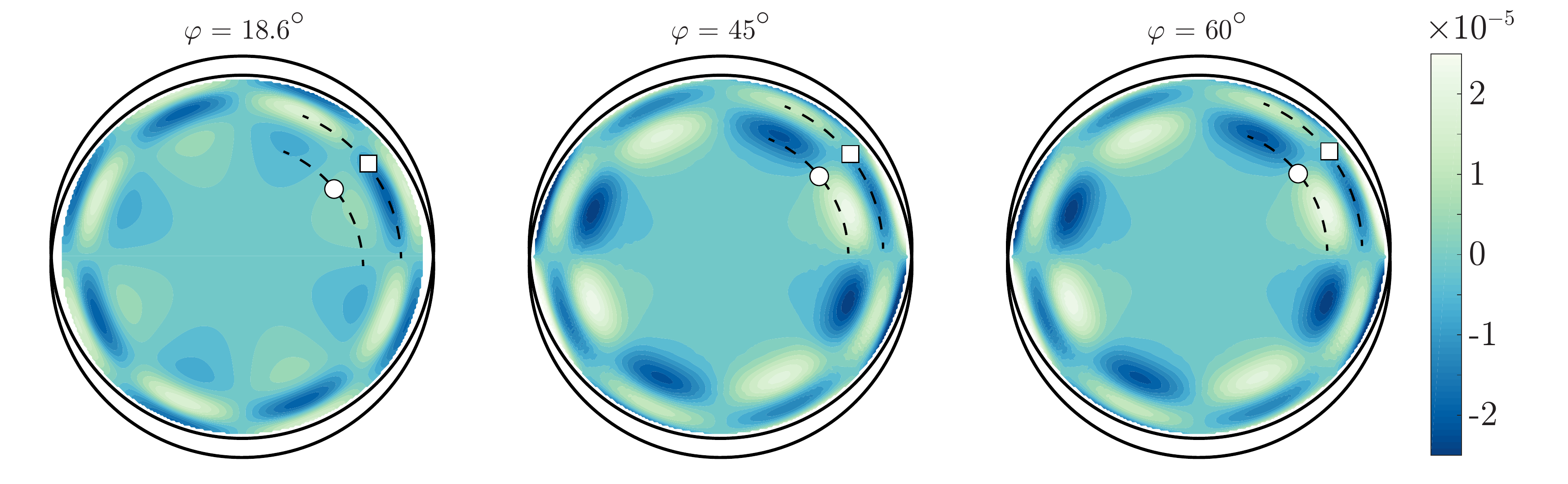}
  \caption{Mean transport of streamwise--averaged vorticity, $(\olbu\cdot\nabla)\olomz$, for $a=0.05, m=1$ and $\Rey_\tau=40$. Dashed curves indicate the inner and outer circles of vorticity production, marked with a circle and square, respectively.}
  \label{fig:prandtlLHS}
\end{figure}

To proceed, we expand all terms in equation \eqref{vorz} in a Fourier--Bessel series 
\be\label{fourierbessel}
  F(r,\theta)= f_0(r) + f_2(r)\sin( 2 m_1\theta)+ f_4(r)\sin(4 m_1\theta)+...
\ee
(cosine terms are zero, and odd terms are prohibited by symmetry) 
where $F$ is any term and $f_0,f_2,f_4...$ are functions. The $\sin(\pm 4 m_1\theta)$ terms largely determine the swirling motion for reasons we now explain.

The mean transport of streamwise--averaged vorticity, $(\olbu\cdot\nabla)\olomz$, is shown in figure \ref{fig:prandtlLHS} for the three different crossing angles, with $a=0.05, m=1$ and $\Rey_\tau=40$. We see that in all cases the amplitude is similar, in the order of $10^{-5}$ in these cases, and the leading contribution is $\propto \sin(4 m \theta)$. We observe that the significant transport of mean streamwise vorticity is organised in a pattern of concentric rings. Consider the two innermost rings in all figures (a thin ring very near the edge is also manifest which does not appear to affect the streamwise--averaged flow perceptibly so we shall ignore this fact). In figure \ref{fig:prandtlLHS} we have indicated the inner and outer rings with a circle and square, respectively. For the two larger angles the two rings have similar amplitudes and comparison with the streamlines in figure \ref{fig:psi}a--c 
shows that the extrema of $(\olbu\cdot\nabla)\olomz$ correspond to oppositely directed rotating motion, that due to the inner ring in the form of elongated streamlines loops confined to an area close to the wall. The larger mean--flow paths correspond to maxima in the inner ring. In comparison the outer ring in the $\varphi=18.6^\circ$ case is similar to the other two cases but for being slightly shifted away from the axis, but strikingly the inner ring is much weaker than the outer, allowing the vortices created by vorticity production in the outer ring to reach into the bulk flow causing mean rotation in the opposite sense. Apparently the presence of Langmuir forcing, instead of adding another source of vorticity production effects a partial cancellation of net inner--ring mean--vorticity production, a conclusion which is surprising to us and should be further investigated in the future. 

\begin{figure}
  \includegraphics[width=\textwidth]{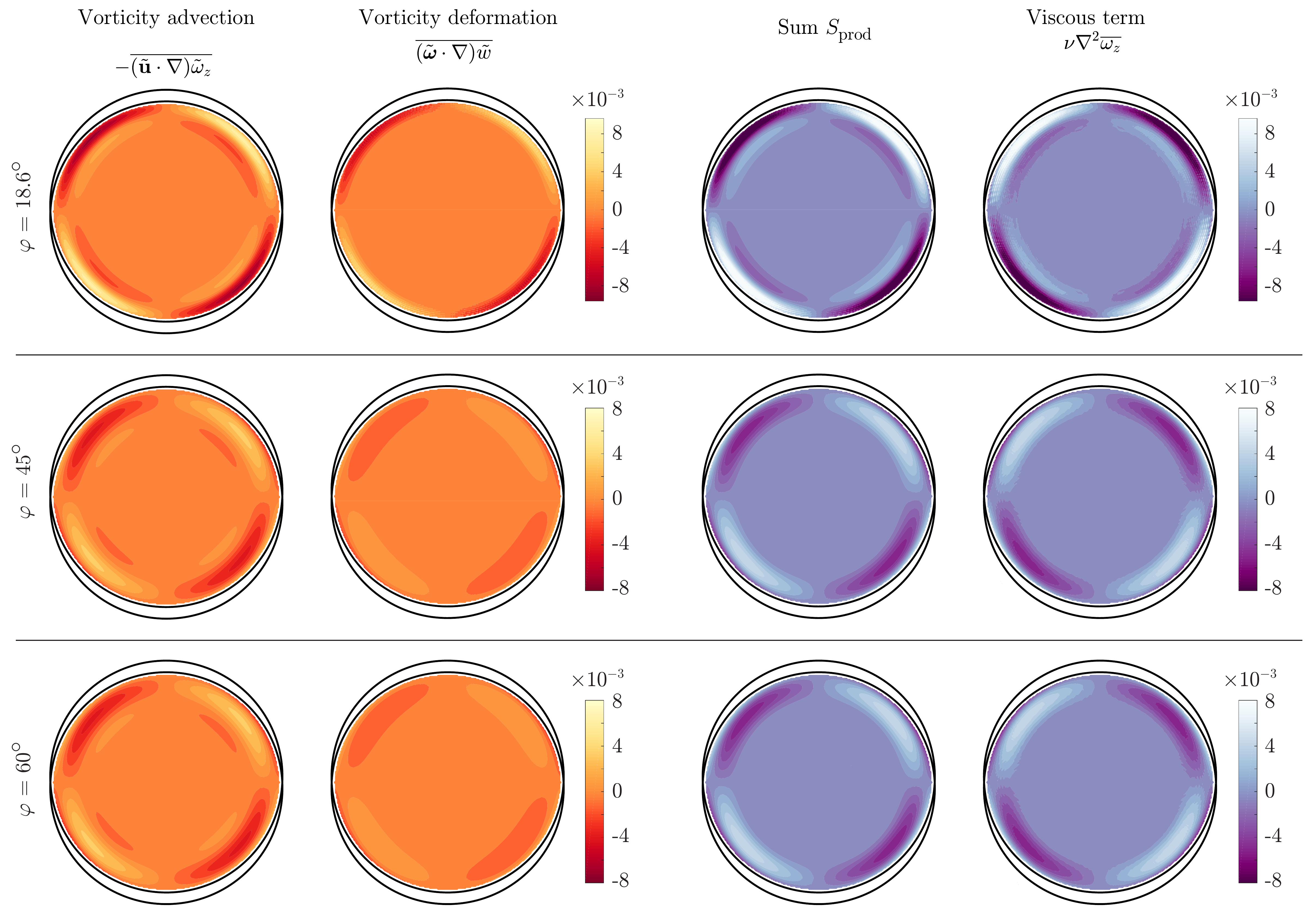}
  \caption{Terms on the right--hand--side of \eqref{vorz}. Each row corresponds to a case in figure \ref{fig:prandtlLHS} with $\varphi$ as indicated.}
  \label{fig:stress_full}
\end{figure}

To continue we analyse the production contributions due to undulatory motion and the viscous diffusion of average streamwise vorticity. 
In figure \ref{fig:stress_full} we have calculated and visualised the terms on the right--hand side of \eqref{vorzb} in full. Comparing with figure \ref{fig:prandtlLHS} we observe that the magnitude of the right--hand terms individually are more than two orders of magnitude larger than those on the left--hand side; the vast majority of mean vorticity production $\Sprod$ is cancelled by viscous diffusion $\omdiss$. The dominant contribution in all panels of \ref{fig:stress_full} is $\propto \sin(2m \theta)$, with higher harmonics only small corrections, and this term cancels in sum, we conjecture, exactly (numerically its amplitude is consistently beneath the level of numerical noise). 

\begin{figure}
  \includegraphics[width=\textwidth]{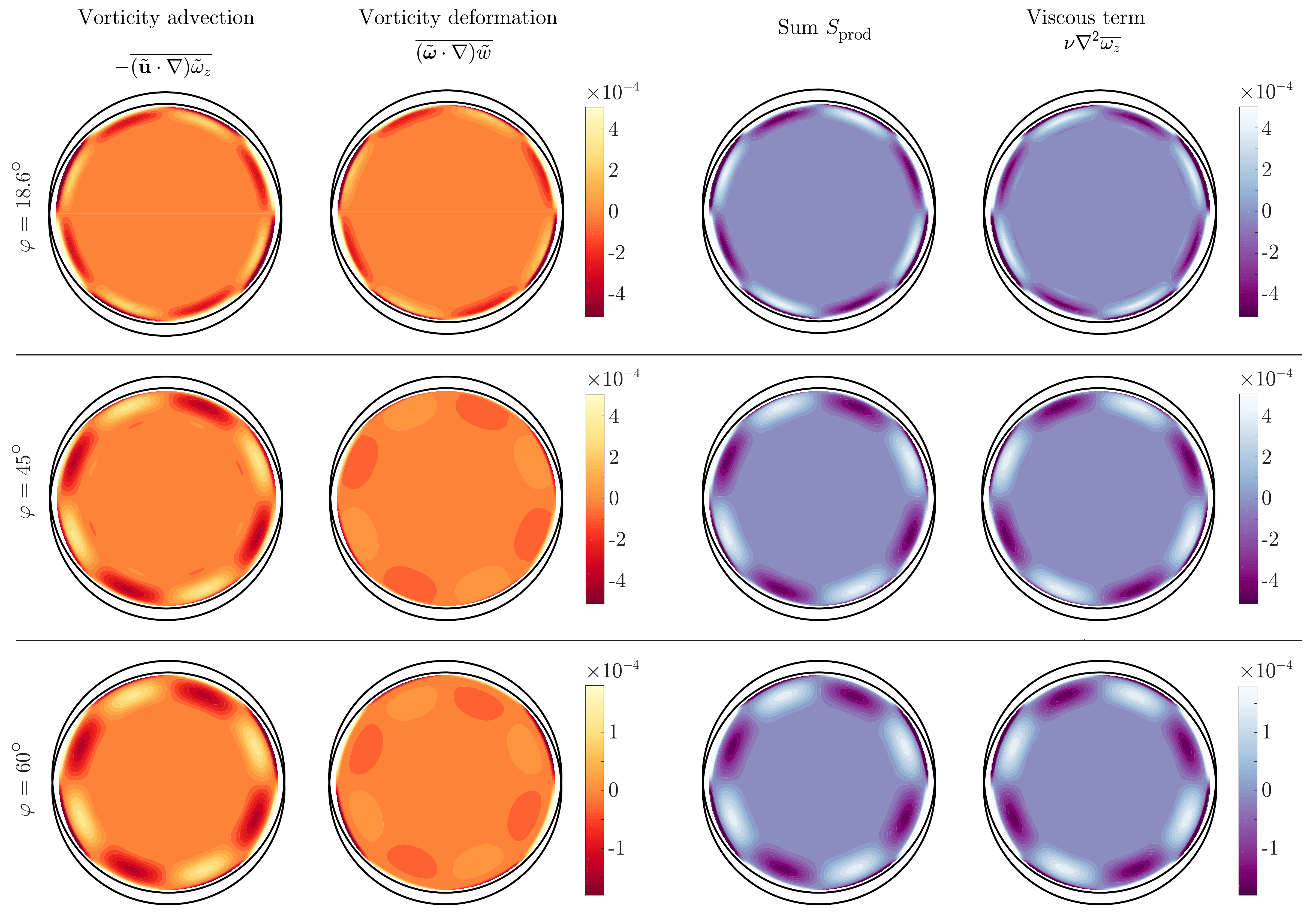}
  \caption{Same as in figure \ref{fig:prandtlLHS}, but the terms $\propto \sin(4 m\theta)$ only.}
  \label{fig:stress_4th}
\end{figure}

Now, it can be observed from the streamwise-averaged flow patterns that the left--hand side of \eqref{vorz} varies no more slowly than $\sim \sin(\pm 4 m \theta)$ as a function of $\theta$. This is in fact a necessity given the observed mean flow patterns in figures \ref{fig:psi} and \ref{fig:pressure} (and all other simulation cases; see Supplementary materials) as we now argue. For concreteness, take the $m=1$ cases in figure \ref{fig:psi} as example. Note that the streamlines are all closed within single quadrants of the cross-section, and consider that the cumulated (integrated) production and diffusion of mean vorticity around a closed streamline must be zero. Since only the $f_4$-term and higher change take both signs inside a single quadrant, periodicity demands $f_0=f_2=0$ for the left-hand side of equation \eqref{vorz}, and hence also for the full right--hand side, as a whole. The argument trivially extends to $m_1>1$, where the same is also invariably observed. 
The $\sin(2m\theta)$ mode which dominates the right--hand--side terms have the same sign and roughly the same magnitude in all cases. While there is some difference in the production due to vortex deformation, it does not seem significant for our purposes given that the $\sin(2m\theta)$ terms do not drive the circulating motions as previously argued. 
We conjecture that the exact cancellation of $\sin(2m\theta)$ terms can be proved in general, but consistent observation in both theory and simulation is sufficient for our purposes.

Since the dominant $\sin(2m\theta)$-mode does not contribute to the net production of mean vorticity, considerably improved clarity is achieved by subtracting it in our plots. Noting that higher harmonics beyond $f_4$ make up only a small correction, we retain only the $f_4$ term which is the main driver of vortical motion. Using
\be
  f_4(r) = \frac2\pi  \int_0^{\pi}\rmd\theta'\, F(r,\theta')\sin (4m\theta'),
\ee 
and plot the same production and diffusion terms again, in figure \ref{fig:stress_4th}. We still observe that the majority of production is cancelled by viscous dissipation --- the amplitudes in figure \ref{fig:stress_4th} are an order of magnitudes higher than those of figure \ref{fig:prandtlLHS}. As previously we again observe that for the same $a, m$ and $\Rey_\tau$ the results for $45^\circ$ and $60^\circ$ are highly similar (but for an overall factor in figure \ref{fig:stress_4th}) while the $18.6^\circ$ is qualitatively different. 

Figure \ref{fig:stress_4th} tells an interesting story. Consider first $45^\circ$ and $60^\circ$. Strong production in the outer ring is due to average advecton of undulating vorticity, most of which is cancelled by the viscous term. A smaller production in the inner ring due to vorticity deformation is evident in figure \ref{fig:prandtlLHS} for these two angles, and these are \emph{not} cancelled by viscous diffusion. The indication is that although weak compared to advection production, due to cancellations this  inner--ring production nevertheless drives the prevailing secondary motion evident in streamline plots e.g.\ in figure \ref{fig:psi} (note the difference in scale between figures \ref{fig:prandtlLHS} and \ref{fig:stress_4th}). 

The picture for $18.6^\circ$ is strikingly different. Here the outer--ring production has the opposite sign, and contributions from advection and deformation are roughly equal, in contrast to the larger angles which for which deformation contributes negligibly to the outer--ring production. 
Crucially the inner--ring, oppositely directed production from deformation is far weaker for $18.6^\circ$, invisible at this scale. 

Naturally we have little basis to predict the extent to which these observations carry over to turbulent flow, although an indication that it could be so is that the production terms on the right--hand side of the turbulent equivalent of equation \eqref{vorz} (with averaging now taken to mean Raynolds averages) are confined to the roughness sublayer where the flow is not strongly turbulent \citep{anderson15}, yet the resulting vortices themselves extend far beyond spanning much of the boundary layer when the roughness is regular in the spanwise direction \citep{willingham14,vanderwel15}. This is a question for future study.

%

}


\section{Summary}
\label{sec:summary}

By furnishing the walls of a pipe flow with a pattern of crossing waves, longitudinal vortices can be made by design through a kinematic mechanism of Langmuir circulation, `CL1', which functions by redirecting the vorticity inherently present in the main flow with no additional forcing. 
The dependence of the vortical secondary motion on Reynolds number $\Rey$, wave crossing angle $\varphi$ and amplitude $a$ was studied with direct numerical simulation throughout the laminar regime, and analysed with a simple theoretical model. The CL1 forcing scales as $\Rey\, a^2$ for small $a$, is strongest for $\varphi \lesssim 30^\circ$ (`contracted eggcarton'), changes sign in the vicinity of $45^\circ$ and is typically oppositely directed and much weaker for $\varphi\gtrsim 45^\circ$ (`protracted eggcarton').
Simulations show how secondary vortices in the opposite sense are also driven by a dynamic mechanism due to differences in wall friction over the wall's crests/troughs vs saddle-points, a mechanism which is present at all $\varphi$. 
For the contracted eggcarton the two effects compete, with CL1 prevailing at small $\varphi \sim 10-20^\circ$ where it is strongest, above which the direction of swirling is reversed. 

When CL1 is negligible, circulation strength scales proportional to $\Rey$ as would be expected. Curiously, for $\varphi=18.6^\circ$ where CL1 is strong, circulation increases significantly faster than $\sim\Rey$. 

Increasing the wall-wave amplitude of the contracted pattern also causes flow reversal, which we attribute to the weakening of CL1 driving due to flow separation. 

\srev{
An analogy exists between streamwise--averaged flow in periodic laminar flow and Prandtl's secondary motion of the second kind in turbulence. In both cases, a transport equation for average streamwise vorticity $\olomz$ is used, and we analyse the source and sink terms and their effect on vortical motion for three different crossing angles, $\varphi=18.6^\circ, 45^\circ$ and $60^\circ$ for $a=0.05, \Rey_\tau=40$. Again the picture is strikingly different for the smallest angle, where Langmuir forcing is strong, to the two larger where it plays a negligible role. In all cases the transport of $\olomz$ is organised in a ring--like structure with extrema in the two rings contribute to rotating flow in opposite senses. For the larger angles the inner ring decides the main swirling motion. For $\varphi=18.6^\circ$ however, the production in the inner ring is far weaker than that in the outer, with vortical motion due to outer--ring production prevailing resulting in flow in the opposite sense. For the larger angles the outer ring (closest to the pipe wall) is mainly driven by advection of vorticity and the inner by deformation (stretching). For  $\varphi=18.6^\circ$ on the other hand, advection and deformation terms contribute approximately equally to production in the outer ring, with the same sign. 
In all cases the vast majority of production of mean streamwise vorticity is cancelled by viscous diffusion, with net production two orders of magnitude smaller than the individual contributions from vorticity advection and deformation.

The effect of imposing the CL1 mechanism on turbulent pipe flow remains an open question for the future. From our observations we conjecture that it could relate to the previous observation by \citet{bhaganagar04} that the crossing angle of small--scale eggcarton rougness has marked effects extending into the outer turbulent boundary layer. 
}

\section*{Acknowledgements}
AHA and S\AA E were funded by the Research Council of Norway (programme FRINATEK), grant number 249740. Supercomputer resources provided by the University of Melbourne and UNINETT Sigma2 --- the National Infrastructure for High Performance Computing and 
Data Storage in Norway. We benefited from discussions with Prof Bruno Eckhardt. Declaration of Interests: The authors report no conflict of interest.

\appendix

\srev{

\section{Initial growth rate of Langmuir vortices}
\label{appendix}

The Orr--Sommerfeld-type boundary value problem, equation \eqref{eq:Rayleigh_O2}, permits an analytical solution describing the initial growth of Langmuir vortices, when the latter are assumed to be inviscid at early times. 
By setting $\Rey^{-1}=0$ one obtains solution 
\begin{equation}
u_r(r,t) = \frac{r t}{2\mtwo} \sum_{s=\pm1}s  \br{
\int_1^0\! 
\dd \rho \,	\frac{\rho^{\mtwo}}{r^{s\mtwo}}-
\int_1^r\! 
\dd \rho \,\frac{\rho^{s\mtwo}}{r^{s\mtwo}} 
}\frac{\rho^2}{r^2} \mc R (\rho),
\label{eq:urO2_sol_inviscid}
\end{equation}
where $\mc R(r)$ given in \eqref{eq:R}. 
The axial velocity component is 
\begin{equation}
u_z(r,t) = -\half t
\U'(r)\, u_r(r,t) \propto a^2 t^2.
\label{eq:uz_inviscid}
\end{equation}
Thus, $u_r$ and $u_\theta$ grow linearly in time whereas $u_z$ grows quadratically.

\bigskip
\begin{center}%
  \begin{tabular}{ll}
  a) & b) \\
  \includegraphics[width=.45\columnwidth]{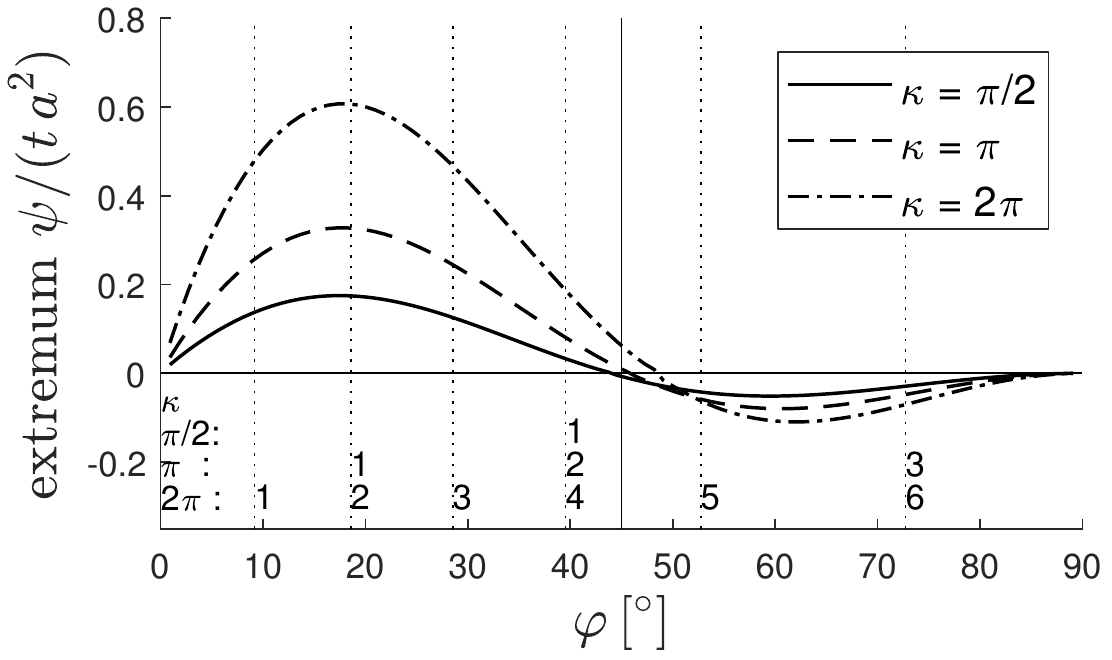} 
  &\includegraphics[width=.45\columnwidth]{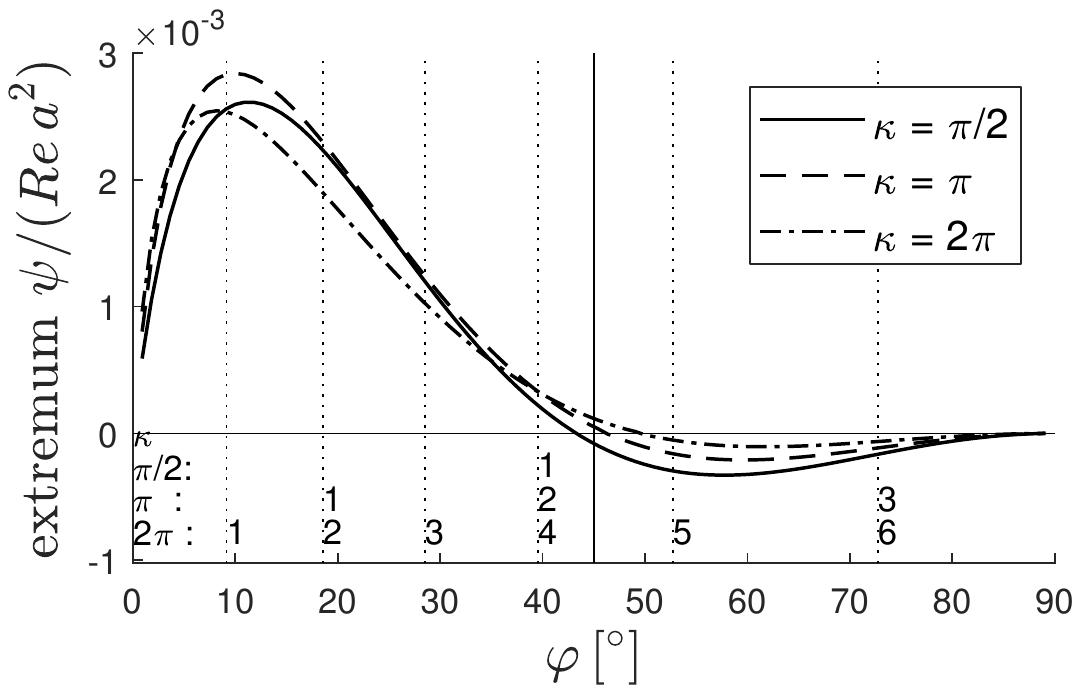}\\  
  c) & \\
  \multicolumn{2}{l}{
    \includegraphics[width=\textwidth]{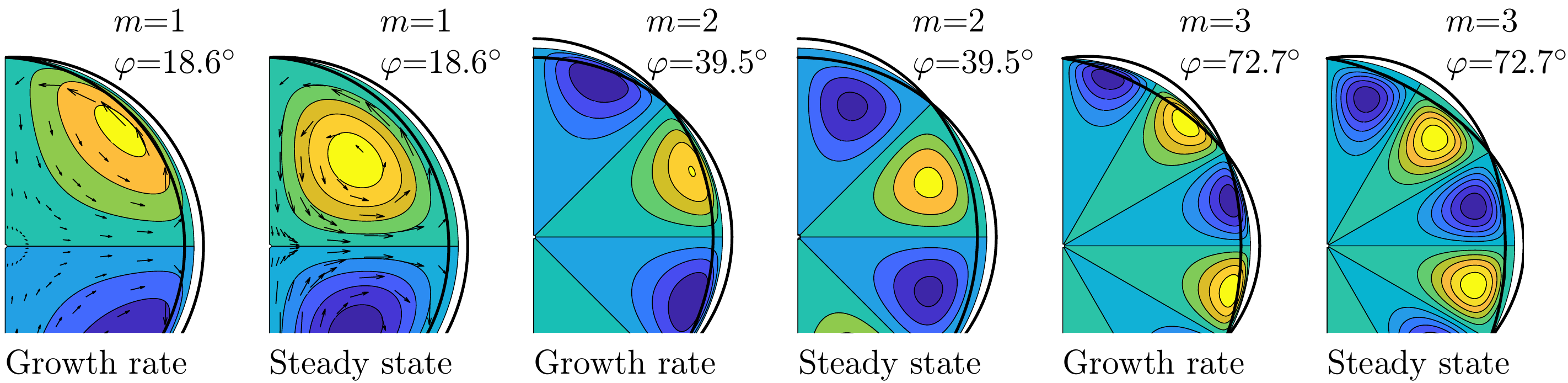}
  }
  \end{tabular}
\captionof{figure}{ Initial growth rate compared to final steady--state solution. a) Initial growth rate, inviscid, b) Ultimate, steady--state solution, c) streamlines.}%
\label{fig:initgrowth}%
\end{center}
\bigskip

A qualitative comparison between solutions of initial growth rate and ultimate state of the CL1-driven vortices are shown in Figure \ref{fig:initgrowth}. Figure \label{fig:init:inviscid} shows dependence on crossing angle $\varphi$ for fixed $\kappa$; compare with steady--state \label{fig:init:viscous} which is the same as Figure \ref{fig:theory}a. Streamlines (contours of $\psi(r,\theta)$) are shown in Figure \ref{fig:initgrowth}c. The second, fourth and sixth panel from the left, labelled `steady state' are the same as Figures \ref{fig:theory}b-d. Notably, vortices move away from the wall after creation before reacing steady state. The same trend was seen theoretically also for flow over a flat plate by \citet{akselsen20}. 

%

}
\bibliographystyle{jfm}
\bibliography{refs_wave}

\end{document}